\documentstyle[epsfig]{article}
\begin{document}
\newcommand{\be}{\begin{equation}}
\newcommand{\ee}{\end{equation}}
\begin{center}
{\Large{\bf Random Coupling of Chaotic Maps leads to Spatiotemporal
Synchronisation}}\\
\vspace{1cm} Sudeshna Sinha\footnote{sudeshna@imsc.ernet.in}\\
{\it Institute of Mathematical Sciences, Taramani, Chennai 600 113,
India}
\end{center}
\begin{abstract}
  We investigate the spatiotemporal dynamics of a network of coupled
  chaotic maps, with varying degrees of randomness in coupling
  connections. While strictly nearest neighbour coupling never allows
  spatiotemporal synchronization in our system, randomly rewiring some
  of those connections stabilises entire networks at $x^*$, where
  $x^*$ is the strongly unstable fixed point solution of the local
  chaotic map. In fact, the smallest degree of randomness in spatial
  connections opens up a window of stability for the synchronised
  fixed point in coupling parameter space. Further, the coupling
  $\epsilon_{bifr}$ at which the onset of spatiotemporal
  synchronisation occurs, scales with the fraction of rewired sites
  $p$ as a power law, for $0.1 < p < 1$. We also show that the
  regularising effect of random connections can be understood from
  stability analysis of the probabilistic evolution equation for the
  system, and approximate analytical expressions for the range and
  $\epsilon_{bifr}$ are obtained.
\end{abstract}

\section{Introduction}

The Coupled Map Lattice (CML) was introduced as a simple model
capturing the essential features of nonlinear dynamics of extended
systems \cite{cml}. Over the past decade research centred around CML
has yielded suggestive conceptual models of spatiotemporal phenomena,
in fields ranging from biology to engineering. In particular, this
class of systems is of considerable interest in modelling phenomena as
diverse as josephson junction arrays, multimode lasers, vortex
dynamics, and even evolutionary biology. The ubiquity of distributed
complex systems has made the CML a focus of sustained research
interest.

A very well-studied coupling form in CMLs is nearest neighbour
coupling. While this regular network is the chosen topology of
innumerable studies, there are strong reasons to re-visit this
fundamental issue in the light of the fact that some degree of
randomness in spatial coupling can be closer to physical reality than
strict nearest neighbour scenarios. In fact many systems of
biological, technological and physical significance are better
described by randomising some fraction of the regular links [2-7]. So
here we will study the spatiotemporal dynamics of CMLs with some of
its coupling connections rewired randomly, i.e. an extended system
comprised of a collection of elemental dynamical units with varying
degrees of randomness in its spatial connections.

Specifically we consider a one-dimensional ring of coupled logistic
maps. The sites are denoted by integers $i = 1, \dots, N$, where $N$
is the linear size of the lattice. On each site is defined a
continuous state variable denoted by $x_n (i)$, which corresponds to
the physical variable of interest. The evolution of this lattice, under
standard nearest neighbour interactions, in discrete time $n$ is given
by
\begin{equation}
x_{n+1} (i) = (1 - \epsilon) f(x_n (i)) + \frac{\epsilon}{2}\{
x_n (i+1) + x_n (i-1) \}
\end{equation}
The strength of coupling is given by $\epsilon$. The local on-site map
is chosen to be the fully chaotic logistic map: $f(x) = 4 x (1 - x)$.
This map has widespread relevance as a prototype of low dimensional
chaos. 

Now we will consider the above system with its coupling connections
rewired randomly in varying degrees, and try to determine what
dynamical properties are significantly affected by the way connections
are made between elements. In our study, at every update we will
connect a fraction $p$ of randomly chosen sites in the lattice, to 2
other random sites, instead of their nearest neighbours as in Eqn.~1.
That is, we will replace a fraction $p$ of nearest neighbour links by
random connections. The case of $p = 0$ corresponds to the usual
nearest neighbour interaction, while $p = 1$, corresponds to
completely random coupling [2-7].  

This scenario is much like small world networks at low $p$, namely $p
\sim 0.01$. Note however that we explore the full range of $p$ here.
In our work $0 \le p \le 1$. So the study is inclusive of, but not
confined to, small world networks.

\section{Numerical Results}

We will now present numerical evidence that random rewiring has a
pronounced effect on spatiotemporal synchronisation. The numerical
results here have been obtained by sampling a large set of random
initial conditions ($\sim 10^4$), and with lattice sizes ranging from
$10$ to $1000$.

Figs.~1 and 2 display the state of the network, $x_n (i) , i= 1, \dots
N$, with respect to coupling strength $\epsilon$, for the limiting
cases of nearest neighbour interactions (i.e. $p = 0$) and completely
random coupling (i.e. $p = 1$). It is clearly seen that the standard
nearest neighbour coupling does not yield a spatiotemporal fixed point
anywhere in the entire coupling range $0 \le \epsilon \le 1$
\cite{fig}. 

Now the effect of introducing some random connections, i.e. $p > 0$,
is to {\it create windows in parameter space where a spatiotemporal
  fixed point state gains stability}, i.e. where one finds all lattice
sites synchronised at $x_n (i) = x^* = 3/4$, for all sites $i$ and at
all times $n$. Note that $x^* = f(x^*)$ is the fixed point solution of
the individual chaotic maps, and is strongly unstable in the local
chaotic map. We then have for all $p > 0$, a stable region of
synchronised fixed points in the parameter interval: $\epsilon_{bifr}
\le \epsilon \le 1.0$. The value of $\epsilon_{bifr}$, where the
spatiotemporally invariant state onsets, is dependent on $p$. It is
evident from Fig.~2 that $\epsilon_{bifr}$ for completely random
coupling $p = 1$ is around $0.62$.

The relationship between the fraction of rewired connections $p$ and
the range, ${\cal R}= (1 - \epsilon_{bifr})$, within which
spatiotemporal homogeniety is obtained, is displayed in Fig.~3. It is
clearly evident that unlike nearest neighbour coupling, random
coupling leads to large parameter regimes of regular homogeneous
behaviour, with all lattice sites synchronised exactly at $x(i) = x^*
= 0.75$. Furthermore the synchronised spatiotemporal fixed point gains
stability over some finite parameter range under {\em any} finite $p$,
i.e. whenever $p > 0$, however small, we have ${\cal R} > 0$. In that
sense strictly nearest neighbour coupling is singular as it does not
support spatiotemporal synchronisation anywhere in coupling parameter
space, whereas any degree of randomness in spatial coupling
connections opens up a synchronised fixed point window. Thus random
connections yield spatiotemporal homogeneity here, while completely
regular connections never do.

The relationship between $\epsilon_{bifr}$, the point of onset of the
spatiotemporal fixed phase, and $p$ is shown in Fig.~4. Note that for
$p < 0.1$ random rewiring does not affect $\epsilon_{bifr}$ much. Only
after $p \sim 0.1$ does $\epsilon_{bifr}$ fall appreciably. Further,
it is clearly evident that for $0.1 < p \le 1$ the lower end of the
stability range falls with increasing $p$ as a well defined power law.
Note that lattice size has very little effect on $\epsilon_{bifr}$,
and the numerically obtained $\epsilon_{bifr}$ for ensembles of
initial random initial conditions over a range of lattice sizes $N =
10, 50, 100$ and $500$ fall quite indistinguishably around each other.

The robust spatiotemporal fixed point supported by random coupling may
have significant ramifications. It has immediate relevance to the
important problem of controlling/synchronising extended chaotic
systems \cite{control, synch}. Obtaining spatiotemporal
synchronisation by introducing some random spatial connections may
have practical utility in the control of large interactive systems.
The regularising effect of random coupling may then help to devise
control methods for spatially extended interactive systems, as well as
suggest natural regularising mechanisms in physical and biological
systems.

\section{Analytical Results}

We shall now analyse this system to account for the much enhanced
stability of the homogeneous phase under random connections. The only
possible solution for a spatiotemporally synchronized state here is
one where all $x_n (i) = x^*$, and $x^* = f(x^*)$ is the fixed point
solution of the local map. For the case of the logistic map $x^* = 4
x^* (1-x^*) = 3/4$.

To calculate the stability of the lattice with all sites at $x^*$ we
will construct an {\it average probabilistic evolution rule} for the
sites, which becomes a sort of {\it mean field version of the
dynamics}.  Some effects due to fluctuations are lost, but as a first
approximation we have found this approach qualitatively right, and
quantitatively close to to the numerical results as well.

We take into account the following: all sites have probability $p$ of
being coupled to random sites, and probability $(1-p)$ of being wired
to nearest neighbours. Then the averaged evolution equation of a site
$j$ is \be x_{n+1}(j) = (1 - \epsilon)f(x_n(j)) + (1-p)
\frac{\epsilon}{2} \left\{ x_n (j+1) + x_n (j-1) \right\} + p
\frac{\epsilon}{2} \left\{ x_n (\xi) + x_n (\eta) \right\} \ee where
$\xi$ and $\eta$ are random integers between $1$ and $N$.

To calculate the stability of the coherent state, we perform the usual
linearization. Replacing $x_n(j) = x^* + h_n(j)$, and expanding to
first order gives

\begin{eqnarray}
h_{n+1}(j) = (1 - \epsilon)f'(x^*)h_{n}(j) &+& (1-p) \frac{\epsilon}{2} \left\{
h_n (j+1) + h_n (j-1) \right\} \\ \nonumber &+& p \frac{\epsilon}{2} \left\{
h_n (\xi) + h_n (\eta) \right\} \\ \nonumber
\approx (1 - \epsilon)f'(x^*)h_{n}(j) &+& (1-p) \frac{\epsilon}{2} \left\{
h_n (j+1) + h_n (j-1) \right\}
\end{eqnarray}
as to a first approximation one can consider the sum over the
fluctuations of the random neighbours to be zero. This approximation
is clearly more valid for small $p$.

For stability considerations one can diagonalize the above expression
using a Fourier transform ($h_n(j) = \sum_{q} \phi_n (q) \exp(i j q)$,
where $q$ is the wavenumber and $j$ is the site index), which
finally leads us to the following growth equation:

\begin{equation}
\frac{\phi_{n+1}}{\phi_n} =  f'(x^*)(1 - \epsilon) + \epsilon (1-p) \cos q
\end{equation}
with $q$ going from $0$ to $\pi$. Clearly the stabilisation condition
will depend on the nature of the local map $f(x)$ through the term
$f^{\prime} (x)$ in Eqn.~4. Considering the fully chaotic logistic map
with $f^{\prime} (x^*) = -2$, one finds that the growth coefficient that
appears in this formula is smaller than one in magnitude if and only
if
\begin{equation}
\frac{1}{1 + p}< \epsilon < 1
\end{equation}
i.e.
\begin{equation}
\epsilon_{bifr} = \frac{1}{1 + p}
\end{equation}
and the range of stability ${\cal R}$ is
\be
{\cal R} = 1 - \frac{1}{1 + p} = \frac{p}{1 + p}
\ee
For small $p$ ($p << 1$) standard expansion gives
\be
{\cal R} \sim p
\ee

The usual case of regular nearest neighbour couplings, $p=0$, gives a
null range, as the upper and lower limits of the range coincide. When
all connections are random, i.e. $p = 1$ the largest stable range is
obtained, and the lower end of the stable window $\epsilon_{bifr}$ is
minimum, with $\epsilon_{bifr} = 1/2$. So stability analysis also
clearly dictates that enhanced stability of the homogeneous phase must
occur under random connections, just as numerical evidence shows.

Fig.~3 exhibits both the analytical expression of Eq.(7) and the
numerically obtained points for comparison. It is clear that for small
$p$ the numerically obtained ${\cal R} \sim p$ is in complete
agreement with the analytical formula. But for higher $p$ some
deviation is discernable, as the ignored effect of the fluctuating
contributions from random neighbours is weighted by $p$, and hence
more pronounced for large $p$. Here the numerically obtained result
goes as \be \epsilon_{bifr} \sim p^{- \phi} \ee for $0.1 < p \le 1$,
with $\phi \sim 0.2$ (see Figs.~4 and 7).

Note that when $p < 0.1$ the effect on $\epsilon_{bifr}$ is not
significant. Only when $0.1 < p \le 1$ does $\epsilon_{bifr}$ fall
appreciably. So connecting elements in a small world network is not
sufficient to make much difference to the onset of the stable
spatiotemporally synchronised state \cite{fx}.

\section{Results from Other Models}

In order to examine the range of applicability of this phenomena we
have examined coupled tent maps and coupled sine circle maps as well.
In the case of coupled tent maps the local map in Eqn.~1 is given as
\be f (x) = 1 - 2 | x - 1/2 | \ee The tent map has an unstable fixed
point at $x^* = 2/3$, with local slope $f^{\prime} (x^*) = -2$.  For
coupled circle map networks the local map in Eqn.~1 is given as \be f
= x + \Omega - \frac{K}{2\pi} \sin (2\pi x) \ee and Eqn.~1 is taken
$mod \ 1$. In the representative example chosen here the parameters of
the circle map are $\Omega = 0, K = 3$.  Here too the local map has a
strongly unstable fixed point at $x^* = \frac{1}{2\pi} \sin^{-1}
(\Omega/K)$, with $f^{\prime} (x^*) = -2$. Numerics very clearly show
that both these systems yield the same phenomena as logistic maps,
namely, one obtains a stable range for spatiotemporal synchronisation
on random rewiring (see Figs.~5 and 6 for the limiting cases of $p =
0$ and $p = 1$ in coupled circle maps).

Since the $f^{\prime} (x^*)$ of both the tent map and the circle map
above is $-2$, we expect from our analysis (Eqn.~4) that their
$\epsilon_{bifr}$ and ${\cal R}$ will be the same as for logistic
maps. This is indeed exactly true, as is evident from Fig.~7 which
displays the point of onset of spatiotemporal synchronisation for all
three cases. In fact the numerically obtained $\epsilon_{bifr}$ values
for ensembles of coupled tent, circle and logistic maps fall
indistinguishably around each other, even for high $p$ where Eqn.~4 is
expected to be less accurate.

Additionally one can infer from the stability analysis above, how
strongly unstable the local maps can possibly be while still allowing
random connections to stabilise the spatiotemporal fixed point. From
Eqn.~4 it follows that the onset of spatiotemporal regularity is
governed by the condition \be | f^{\prime} (x^*) | < \frac{1 -
  \epsilon + \epsilon p}{1 - \epsilon} = 1 + \frac{\epsilon p}{1 -
  \epsilon}\ee Clearly then, locally unstable maps with $|f^{\prime}
(x^*)| > 1$ can be stabilised by any finite $p$, i.e. by any degree of
randomness in the coupling connections. As coupling strength
$\epsilon$ and fraction of rewiring $p$ increases, maps with
increasingly unstable fixed points can be synchronised stably by
random rewiring. For regular coupling ($p = 0$) on the other hand the
local instability can only be as large as $1$, i.e. if and only if the
local components possess stable fixed points can their network be
stabilised at spatiotemporal fixed points, as numerics have already
shown.

Lastly, in certain contexts, especially neuronal scenarios, the
randomness in coupling may be {\em static}. In the presence of such
quenched randomness in the couplings, once again one obtains a stable
range  ${\cal R}$ for spatiotemporal synchronisation. But unlike
dynamical rewiring where the ${\cal R}$ is independent of the size and
initial preparation of the lattice and its connections, here there is
a spread in the values of ${\cal R}$ obtained from different (static)
realisations of the random connections. Futhermore, this distribution
of ${\cal R}$ is dependent on the size of network. For instance, on an
average, networks of size $N = 10$ with fully random static
connections ($p = 1$) yield $\epsilon_{bifr} \sim 0.75$ and those of
size $N = 100$ yield $\epsilon_{bifr} \sim 0.85$, as opposed to
$\epsilon_{bifr} \sim 0.62$ obtained for all $N$ for dynamically
updated random connections.  Fig.~8 displays the average range $<
{\cal R} >$ with respect to network size $N$, indicative of clear
scaling behaviour: \be <{\cal R}> \sim N^{- \nu} \ee with $\nu \sim
0.24$.  This suggests that the range narrows {\em slowly} with
increasing network size. So, while in the limit of infinite lattices
there will be no spatiotemporal synchronisation, for finite networks
static randomness will lead to stable windows of spatiotemporal
synchronisation.

This behaviour can be understood by a examining the linear stability
of the homogeneous solution: $x_n (j) = x^*$ for all sites $j$ at all
times $n$.  For instance for the case of fully random static
connections $p = 1$, considering the dynamics of small perturbations
over the network one obtains the transfer matrix connecting the
perturbation vectors at successive times to be a sum of a $N \times N$
diagonal matrix, with entries $( 1 - \epsilon ) f^{\prime} (x^*)$, and
$\epsilon/2 \times {\bf C}$ where ${\bf C}$ is a $N \times N$ sparse
non-symmetric matrix with two random entries of $1$ on each row. Now
the minimum of the real part of the eigenvalues of ${\bf C}$,
$\lambda_{\rm{min}}$, crucially governs the stability. Typically
$\epsilon_{bifr} = 2/\{ \lambda_{\rm{min}} + 4 \}$ when
$f^{\prime}(x^*) = -2$. Now the values of $\lambda_{\rm{min}}$
obtained from different static realisations of the connectivity matrix
${\bf C}$ are distributed differently for different sizes $N$. For
small $N$ this distribution is broad and has less negative averages
($\sim -1$). On the other hand for large $N$ the distribution gets
narrower and tends towards the limiting value of $-2$. This results in
a larger range of stability, and greater spread in $\epsilon_{bifr}$
for small networks. In fact for small $N$, certain static realisations
yield a larger range of stability (${\cal R } \sim 1/2$) than dynamic
rewiring.

\section{Conclusions}

In summary then, we have shown that random rewiring of spatial
connections has a pronounced effect on spatiotemporal synchronisation.
In fact strictly nearest neighbour coupling is not generic, in that it
does not support any spatiotemporal fixed point phase, while the
smallest degree of random rewiring has the effect of creating a window
of spatiotemporal invariance in coupling parameter space. Further the
regularising effect of random connections can be understood from
stability analysis of the probabilistic evolution equation for the
system, and approximate analytical expressions for the range and onset
of spatiotemporal synchronisation have been obtained. The key
observation that {\it random coupling regularises} may then help to
devise control methods for spatially extended interactive systems, as
well as suggest
natural regularising mechanisms in physical and biological systems.\\

{\em Acknowledgements:} I would like to thank Neelima Gupte for many
stimulating discussions and ideas on the subject.\\

\newpage
\begin{figure}
\mbox{\epsfig{file=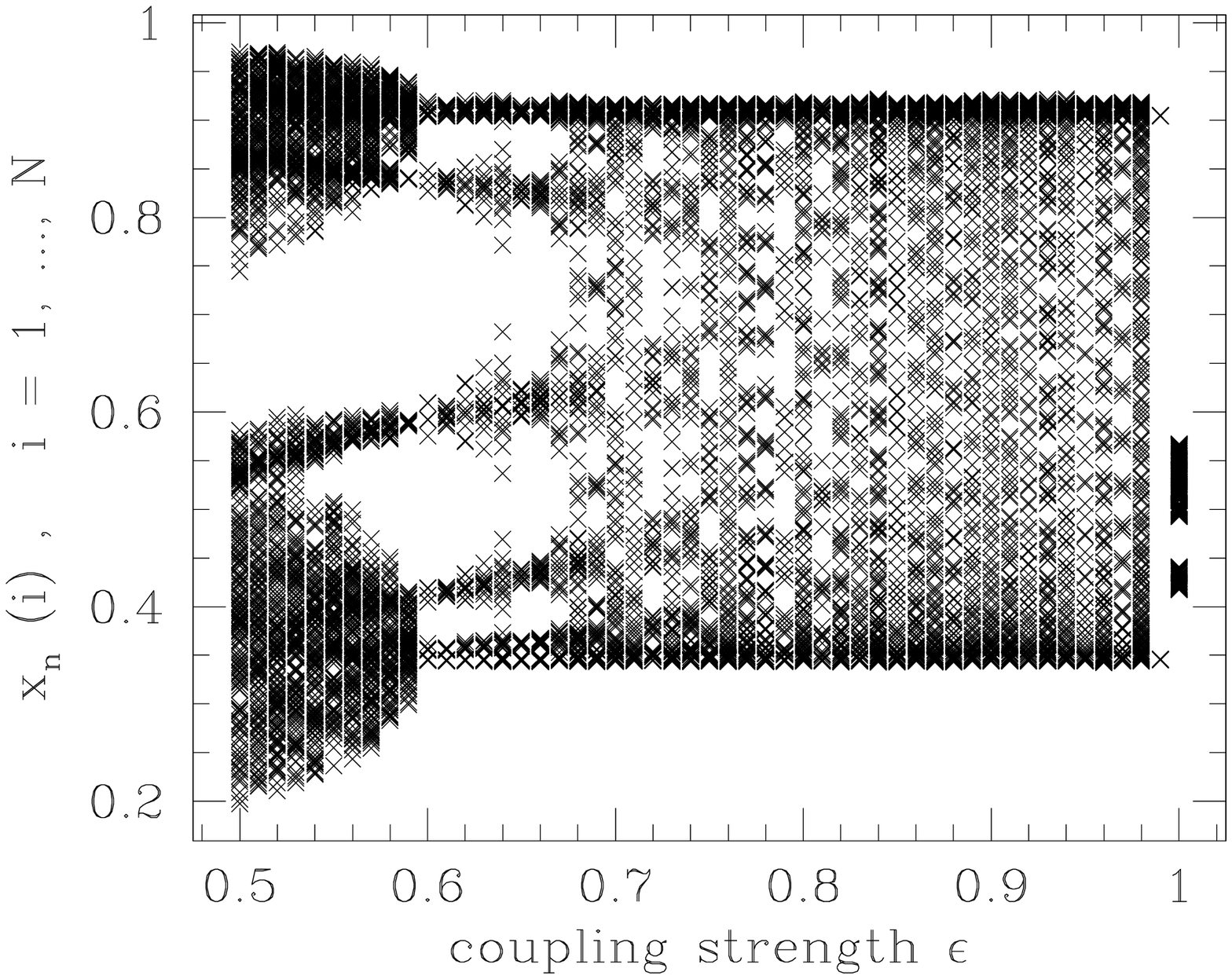,width=12truecm}}
\caption{ Bifurcation diagram showing values of $x_n (i)$ with respect
  to coupling strength $\epsilon$, for coupled logistic maps with
  strictly regular nearest neighbour connections. Here the linear size
  of the lattice is $N = 100$ and in the figure we plot $x_n(i)$ ($i =
  1, \dots, 100$) over $n = 1, \dots 5$ iterations (after a
  transience time of $1000$) for 5 different initial conditions.}
\end{figure}

\begin{figure}
\mbox{\epsfig{file=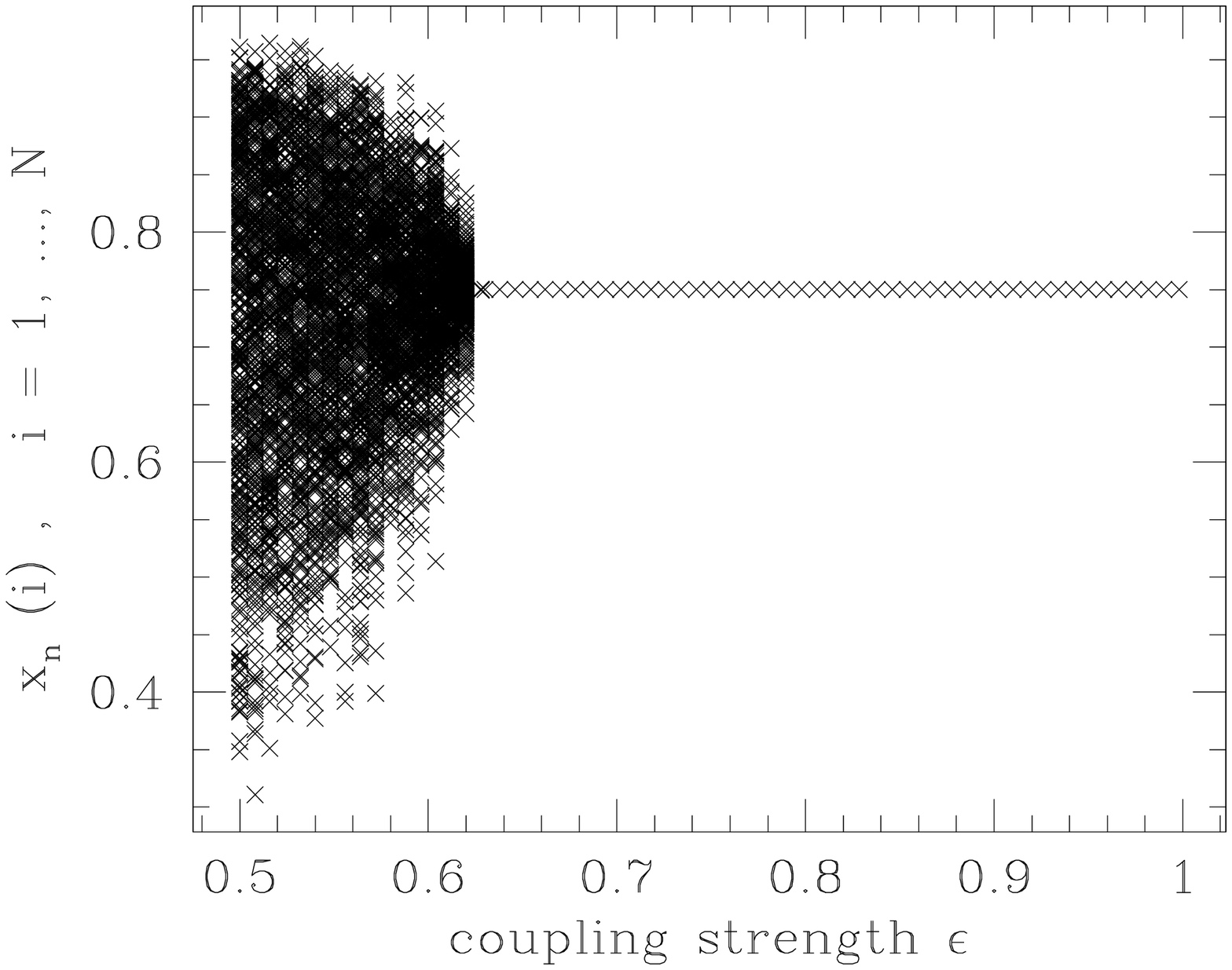,width=12truecm}}
\caption{ Bifurcation diagram showing values of $x_n (i)$ with respect
  to coupling strength $\epsilon$, for coupled logistic maps with
  completely random connections. Here the linear size of the lattice
  is $N = 100$ and in the figure we plot $x_n(i)$ ($i = 1, \dots,
  100$) over $n = 1, \dots 5$ iterations (after a transience time of
  $1000$) for 5 different initial conditions.}
\end{figure}

\begin{figure}
\mbox{\epsfig{file=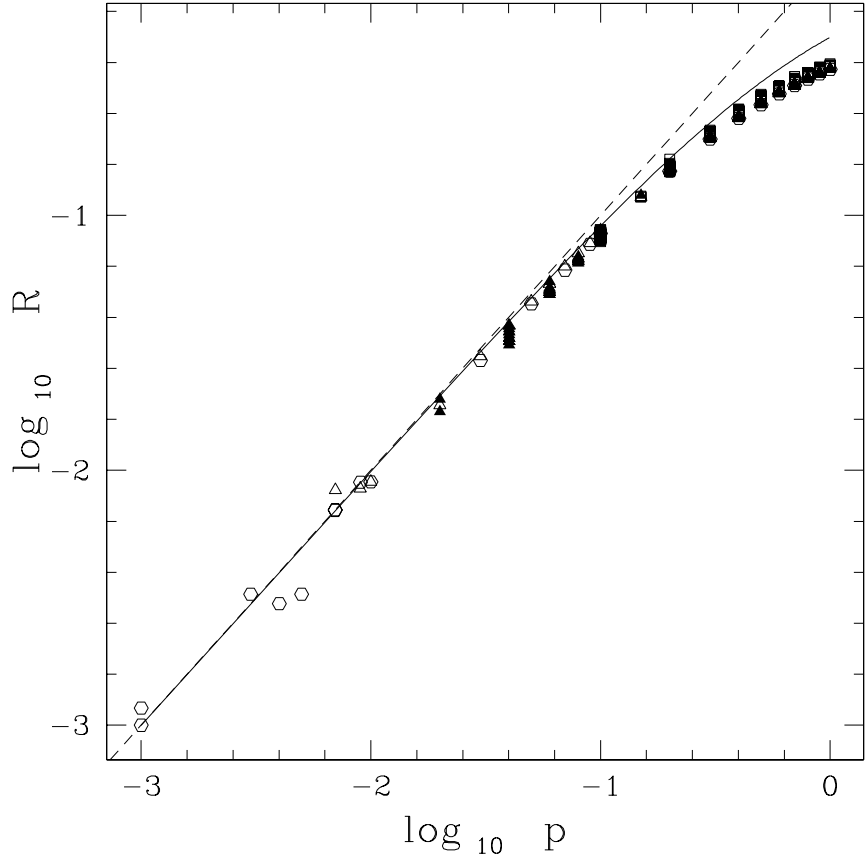,width=12truecm}}
\caption{The stable range {\cal R} with respect to the fraction of
  randomly rewired sites $p$ $(0.001 \le p \le 1)$ : the solid line
  displays the analytical result of Eqn.~7, and the different points
  are obtained from numerical simulations over several different
  initial conditions, for 4 different lattice sizes, namely $N = 10,
  50, 100$ and $500$. The dotted line shows ${\cal R} = p$, and it is
  clear that for a large range of $p$ the approximation holds.}
\end{figure}

\begin{figure}
\mbox{\epsfig{file=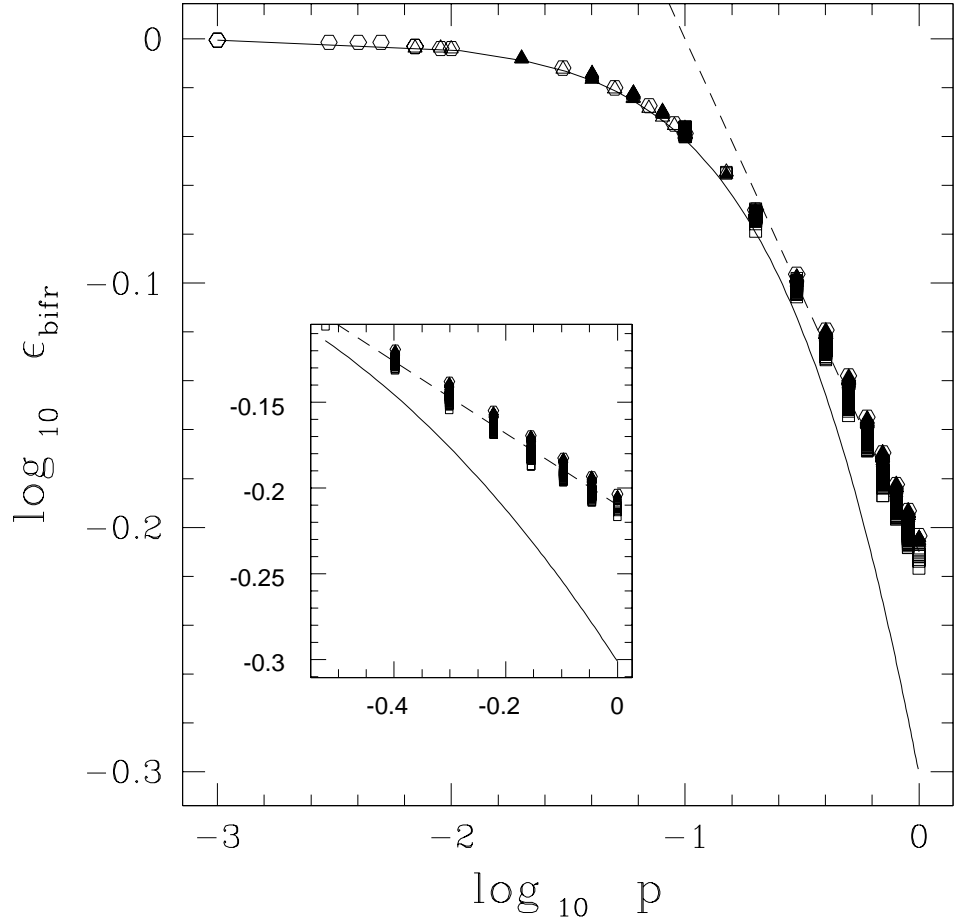,width=14truecm}}
\caption{The $\epsilon_{bifr}$ (i.e. the value of coupling at which
  the onset of spatiotemporal synchronization occurs) with respect to
  fraction of randomly rewired sites $p$ $(0.001 \le p \le 1)$ : the
  solid line displays the analytical result of Eqn.~6, and the points
  are obtained from numerical simulations over several different
  initial conditions, for 4 different lattice sizes, namely $N = 10,
  50, 100$ and $500$. The inset box shows a blow-up of $0.1 < p \le
  1$. Here the numerically obtained $\epsilon_{bifr}$ deviates from
  the mean field results. The dashed line is the best fit straight
  line for the numerically obtained points in that region.}
\end{figure}

\begin{figure}
\mbox{\epsfig{file=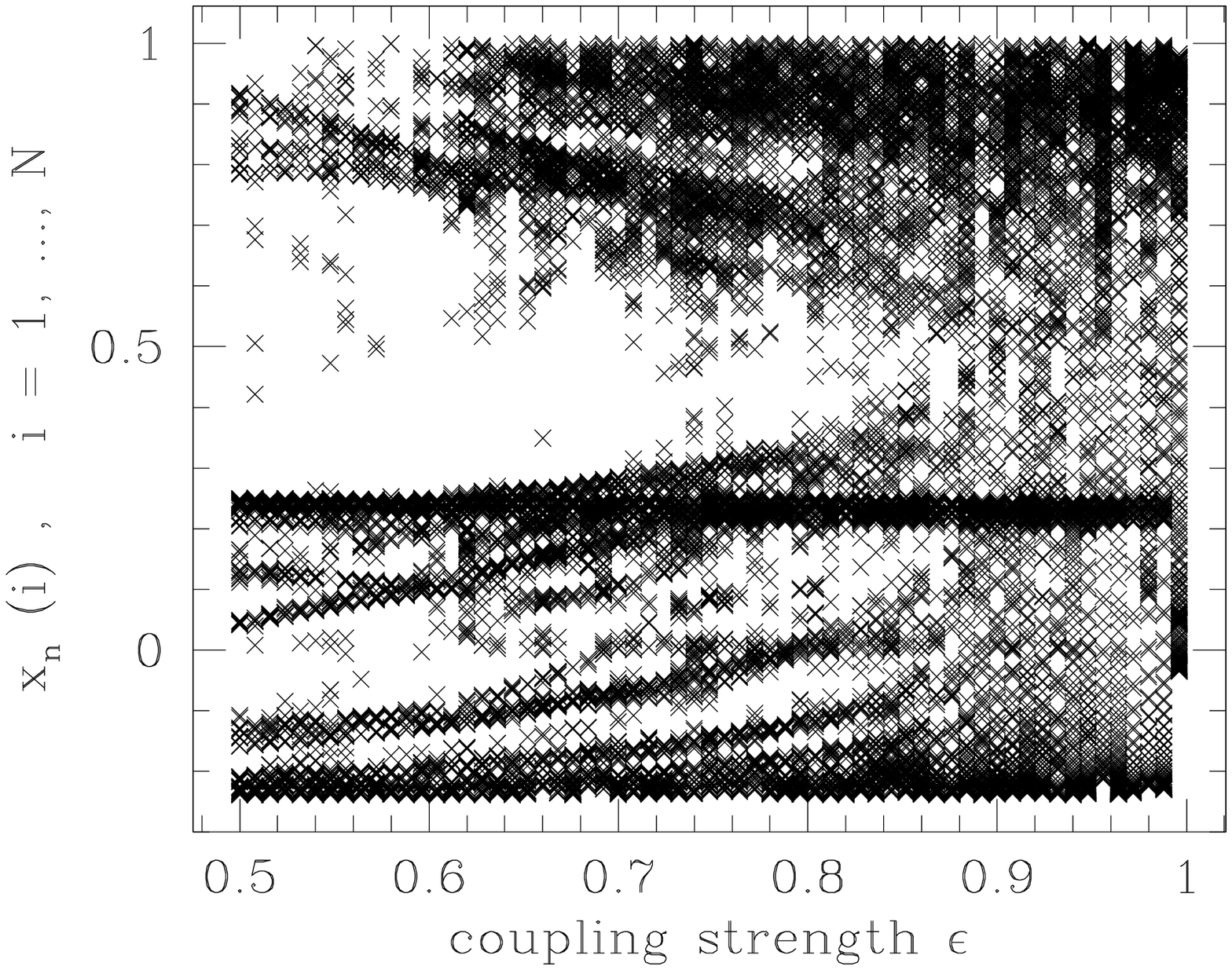,width=12truecm}}
\caption{ Bifurcation diagram showing values of $x_n (i)$ with respect
  to coupling strength $\epsilon$, for coupled sine circle maps with
  strictly regular nearest neighbour connections. Here the linear size
  of the lattice is $N = 100$ and in the figure we plot $x_n(i)$ ($i =
  1, \dots, 100$) over $n = 1, \dots 5$ iterations (after a
  transience time of $1000$) for 5 different initial conditions.}
\end{figure}

\begin{figure}
\mbox{\epsfig{file=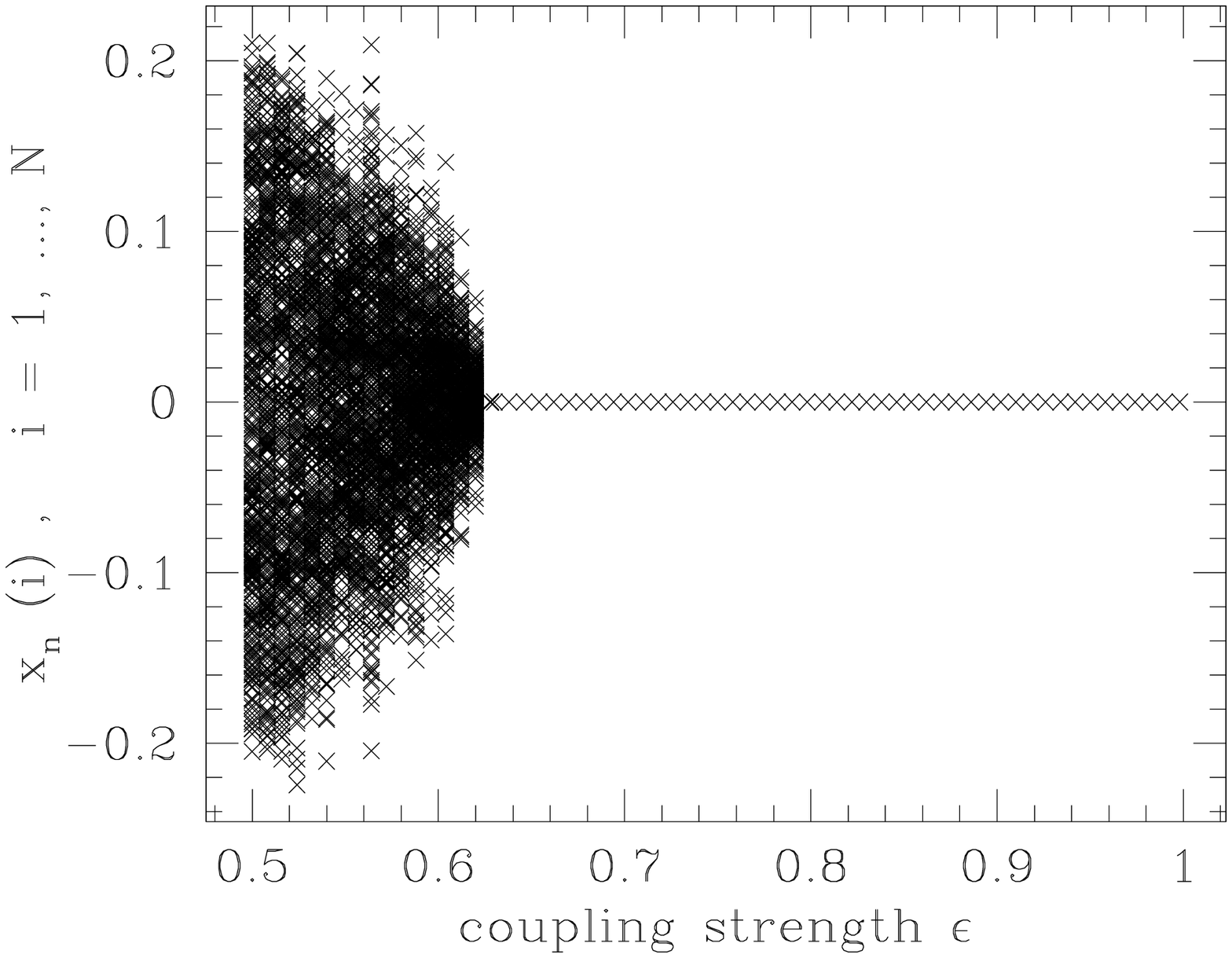,width=12truecm}}
\caption{ Bifurcation diagram showing values of $x_n (i)$ with respect
  to coupling strength $\epsilon$, for coupled sine circle maps with
  completely random connections. Here the linear size of the lattice
  is $N = 100$ and in the figure we plot $x_n(i)$ ($i = 1, \dots,
  100$) over $n = 1, \dots 5$ iterations (after a transience time of
  $1000$) for 5 different initial conditions.}
\end{figure}

\begin{figure}
\mbox{\epsfig{file=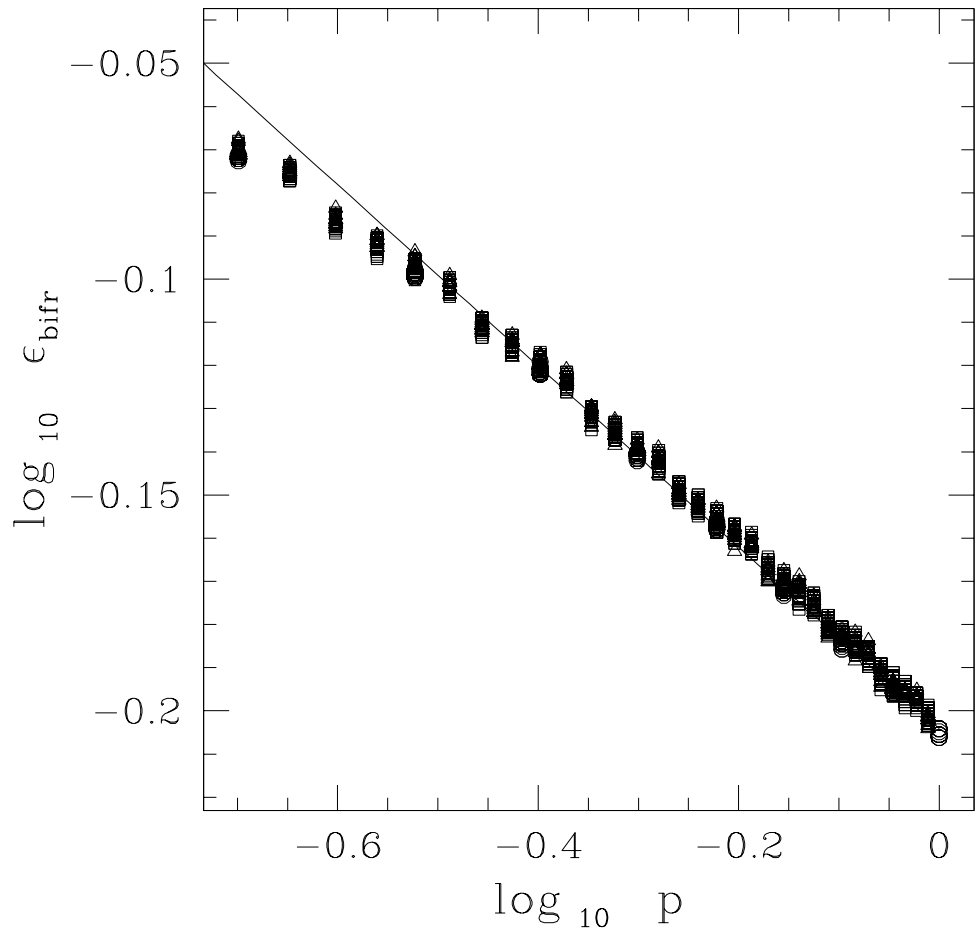,width=14truecm}}
\caption{Plot of $\epsilon_{bifr}$ (i.e. the value of coupling at
  which the onset of spatiotemporal synchronization occurs) with
  respect to fraction of randomly rewired sites $p$ $(0.2 \le p \le
  1)$. The points are obtained from numerical simulations over several
  different initial conditions, for lattice size $N = 50$, for the
  case of (a) coupled tent maps (open squares) (b) coupled circle maps
  (open triangles) and (c) coupled logistic maps (open circles). The
  solid line displays the best fit straight line for the numerically
  obtained points.}
\end{figure}

\begin{figure}
\mbox{\epsfig{file=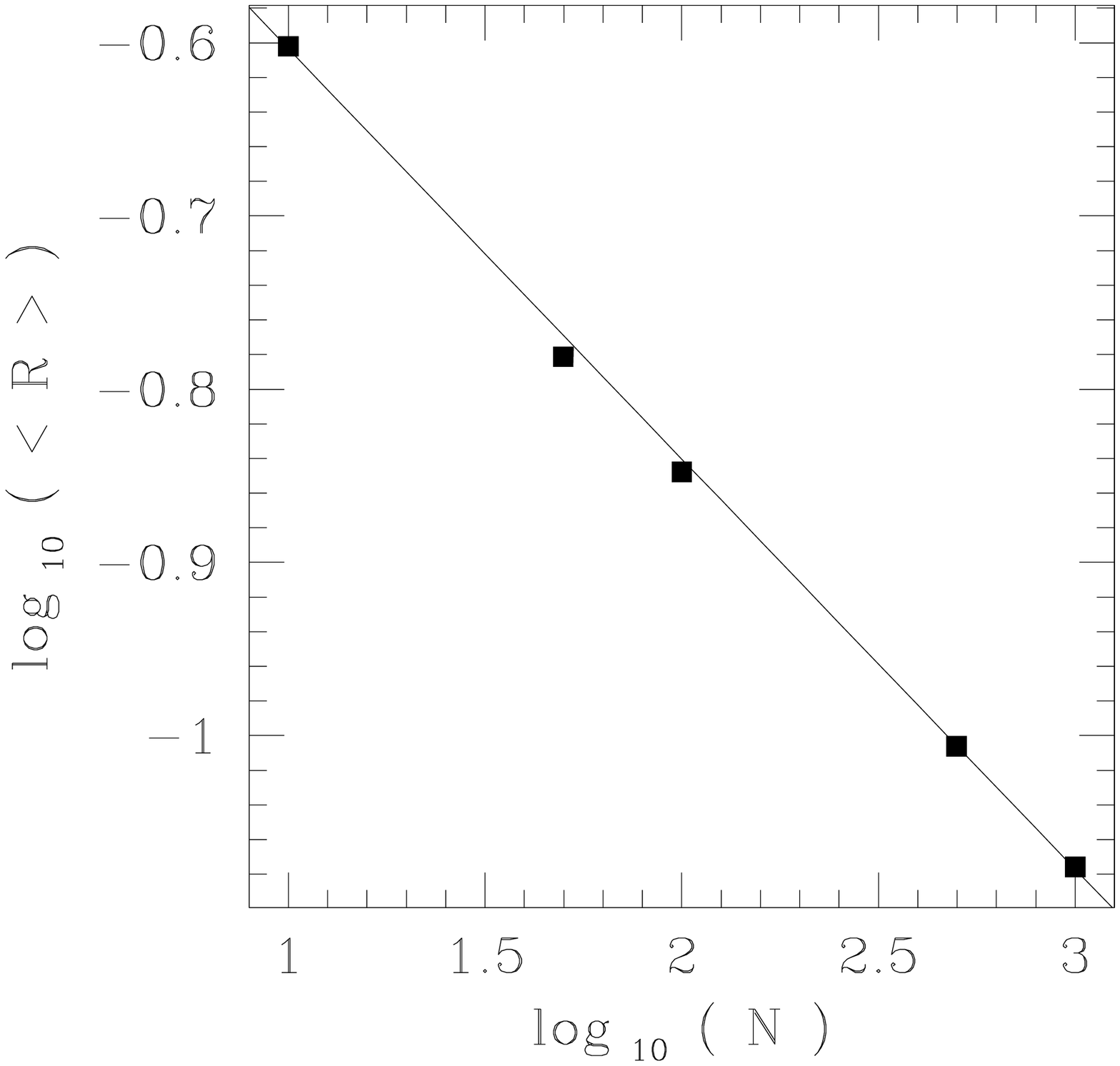,width=12truecm}}
\caption{Plot of the average stable range $< {\cal R} >$ of
  spatiotemporal synchronisation obtained in the case of static random
  connections with respect to network size $N$, for rewired fraction
  $p = 1$.  Here we average ${\cal R}$ over $10^4$ different
  realisations of static random connections. The solid line shows the
  best fit line to the numerically obtained data, indicating clear
  scaling.}
\end{figure}


\begin{thebibliography}{99}
\bibitem{fig} Only the strong coupling range $1/2 \le \epsilon \le 1$
  is displayed in Figs.~1 and 2. The weak coupling range $0 \le
  \epsilon \le 1/2$ (not shown) is strongly irregular for both $p = 0$
  and $p = 1$.
\bibitem{fx} The range of stability also depends on the specific
coupling form: for instance if one considered the future-coupled case,
where the coupling occurs through $f(x)$ then an additional factor of
$f^{\prime} (x)$ multiplies the second term in Eqn.~4. The validity of
the approximation in Eqn.~3 also depends on the coupling form. Again
for the future-coupled case the approximation is much worse as the
term ignored is weighted by an additional factor of $f^{\prime} (x)$,
which is greater than $1$ for chaotic maps. In fact for the
future-coupled case numerical evidence of spatial synchronisation is
obtained for chaotic logistic maps with $f(x) = rx (1-x)$, only when
$r < 4$, e.g. around $r \sim 3.8$. But even for $r \sim 4$, where
exact synchronisation is not obtained, the roughness of the spatial
profile is much less than that for regular nearest neighbour
coupling. So the random coupling does have a regularising effect there
as well, though not as complete as in the examples in this
work.
\end{thebibliography}
\end{document}